\documentclass[aps,prl,twocolumn,superscriptaddress,showpacs]{revtex4}
\usepackage{graphicx}
\usepackage{latexsym}
\usepackage{amssymb}
\usepackage{amsmath}
\usepackage{amsfonts}
\usepackage{bm}
\usepackage{multirow}
\usepackage{color}
\usepackage{comment}
\usepackage{feyn}
\newcommand{\ii}{\mathrm{i}}

\renewcommand{\O}{\mathrm{O}}
\newcommand{\SU}{\mathrm{SU}}
\newcommand{\U}{\mathrm{U}}

\newcommand{\figref}[1]{Fig.\,\ref{#1}}

\newcommand{\beq}{\begin{equation}}
\newcommand{\eeq}{\end{equation}}
\newcommand{\beqn}{\begin{eqnarray}}
\newcommand{\eeqn}{\end{eqnarray}}

\DeclareMathAlphabet{\mathbbold}{U}{bbold}{m}{n}

\def\SU{{\rm SU}}

\def\U{{\rm U}}

\begin{document}

\title{Non-Landau Quantum Phase Transitions and nearly-Marginal non-Fermi Liquid}

\author{Yichen Xu}
\affiliation{Department of Physics, University of California,
Santa Barbara, CA 93106, USA}

\author{Hao Geng}
\affiliation{Department of Physics, University of Washington,
Seattle, WA 98195, USA}

\author{Xiao-Chuan Wu}
\affiliation{Department of Physics, University of California,
Santa Barbara, CA 93106, USA}

\author{Chao-Ming Jian}
\affiliation{Station Q, Microsoft, Santa Barbara, California
93106-6105, USA}

\author{Cenke Xu}
\affiliation{Department of Physics, University of California,
Santa Barbara, CA 93106, USA}

\begin{abstract}
Non-fermi liquid and unconventional quantum critical points (QCP)
with strong fractionalization are two exceptional phenomena beyond
the classic condensed matter doctrines, both of which could occur
in strongly interacting quantum many-body systems. This work
demonstrates that using a controlled method one can construct a
non-Fermi liquid within a considerable energy window based on the
unique physics of unconventional QCPs. We will focus on the
``nearly-marginal non-Fermi liquid", defined as a state whose
fermion self-energy scales as $\Sigma_f(\ii \omega) \sim \ii
\mathrm{sgn}(\omega)|\omega|^{\alpha}$ with $\alpha$ close to $1$
in a considerable energy window. The nearly-marginal non-fermi
liquid is obtained by coupling an electron fermi surface to
unconventional QCPs that are beyond the Landau's paradigm. This
mechanism relies on the observation that the anomalous dimension
$\eta$ of the order parameter of these unconventional QCPs can be
close to $1$, which is significantly larger than conventional
Landau phase transitions, for example the Wilson-Fisher fixed
points. The fact that $\eta \sim 1$ justifies a perturbative
renormalization group calculation proposed earlier. Various
candidate QCPs that meet this desired condition are proposed.

\end{abstract}

\maketitle

\section{Introduction}


In the past few decades, a consensus has been gradually reached
that quantum many-body physics with strong quantum entanglement
can be much richer than classical physics driven by thermal
fluctuations~\cite{wenreviewtopo,wenscience}. Classical phase
transitions usually happen between a disordered phase with high
symmetries, and an ordered phase which spontaneously breaks such
symmetries. Typical classical phase transitions can be well
described by the Landau's paradigm, but the Landau's paradigm may
or may not apply to quantum phase transitions that happen at zero
temperature. Generally speaking, the Landau's formalism can only
describe the quantum phase transition between a direct-product
quantum disordered state and a spontaneous symmetry breaking
state; but it can no longer describe the quantum phase transition
between two states when at least one of the states cannot be
adiabatically connected to a direct product states, $i.e.$ when
this state is a topological order~\cite{wen-wu}; nor can the
Landau's paradigm describe generic continuous quantum phase
transitions between states with different spontaneous symmetry
breakings~\cite{deconfine1,deconfine2,xutriangle}.

Phenomenologically, in contrast with the ordinary Landau's
transitions, non-Landau transitions often have a large anomalous
dimension of order parameters, due to fractionalization or
deconfinement of the order
parameter~\cite{deconfinesandvik1,deconfinesandvik2,deconfinemelko,deconfinedualnumeric}.
The ordinary Wilson-Fisher (WF) fixed point in $(2+1)d$ space-time
(or three dimensional classical space) has very small anomalous
dimensions~\cite{vicari}, meaning that the Wilson-Fisher fixed
point is not far from the mean field theory. In particular, in the
large$-N$ limit, the anomalous dimension of the vector order
parameter of the $\O(N)$ Wilson-Fisher fixed point is $\eta \sim
0$; while the CP$^{N-1}$ model, the theory that describes a class
of non-Landau quantum phase
transition~\cite{deconfine1,deconfine2}, has $\eta \sim 1$ in the
large-$N$ limit~\cite{kaulsachdev}. Numerically it was also
confirmed that the quantum phase transition between the $Z_2$
topological order and the superfluid phase has $\eta \sim
1.5$~\cite{melko1,melko2}, as was predicted theoretically. The
large anomalous dimension has been used as a strong signature when
searching for unconventional QCPs numerically.

In this work we propose that the unique physics described above
about the unconventional QCPs with strong fractionalization can be
used to construct another broadly observed phenomenon beyond the
classic Landau's theory: the non-Fermi liquid whose fermion
self-energy scales $\Sigma_f(\ii \omega) \sim \ii
\mathrm{sgn}(\omega)|\omega|^{\alpha}$ with $\alpha < 1$. When
$\alpha = 1$, this non-fermi liquid is referred to as marginal
fermi liquid~\cite{mfl}. Signature of marginal fermi liquid and
nearly-marginal fermi liquid have been observed rather broadly in
various materials~\cite{review,moirestrange,youngstrange}. In this
work we will focus on the non-Fermi liquid that is
``nearly-marginal", meaning $\alpha$ is close to $1$.

We assume that there exists a field $\mathcal{O}(\boldsymbol{x},
\tau)$ in the unconventional QCP that carries zero momentum, and
it couples to the fermi surface in the standard way: $\int d^2x
d\tau \ g \psi^\dagger T \psi \mathcal{O}$, where $T$ is a flavor
matrix of the fermion. We assume that we first solve (or
approximately solve) the bosonic part of the theory, $i.e.$ the
strongly interacting QCP without coupling to the fermi surface,
and calculate the anomalous dimension $\eta$ at the QCP: \beqn
\langle \mathcal{O}(\boldsymbol{q}, \omega)
\mathcal{O}(-\boldsymbol{q},-\omega) \rangle \sim
\frac{1}{\Omega^{2 - \eta}} \eeqn where $\Omega \sim \sqrt{v^2
\boldsymbol{q}^2 + \omega^2}$. Then the fermion self-energy, the
quantity of central interest to us, is computed perturbatively
with the boson-fermion coupling $g$.

When the anomalous dimension $\eta$ is close to $1$, we can take
$\eta = 1 - \epsilon$ with small $\epsilon$.
Ref.~\cite{nayak1,nayak2,mrossnfl} developed a formalism for the
boson-fermion coupled theory with an expansion of $\epsilon$,
though eventually one needs to extrapolate the calculation to
$\epsilon = 1$ for the problems studied
therein~\cite{nayak1,nayak2,mrossnfl}, and the convergence of the
$\epsilon-$expansion at $\epsilon = 1$ is unknown, $i.e.$ even if
we start with a weak boson-fermion coupling, it would become
nonperturbative under renormalization group (RG). But we will
demonstrate in the next section that in the cases that we are
interested in, $\epsilon$ is naturally small when $\eta$ is close
to $1$, due to the fractionalized nature of many unconventional
QCPs. To the leading nontrivial order, our problem can be
naturally studied by the previously proposed perturbative
formalism with small $\epsilon$.

Here we stress that our goal is to construct a scenario in which a
non-Fermi liquid state within an energy window can be constructed
using a controlled method. Recently many works have taken a
similar spirit, and various non-Fermi liquid states especially a
state that mimics the strange metal were constructed by deforming
the soluble Sachdev-Ye-Kitaev (SYK) and related
models~\cite{SachdevYe1993,Kitaev2015,MaldacenaStanford2016,Witten2016,Klebanov2016}.
Then within the energy window where the deformation remains
perturbative, the system resembles the non-Fermi
liquid~\cite{Song2017,patel2017,patel2018,berg2018,xu2018,xu2019}.
Our current work also starts with (approximately) soluble strongly
interacting bosonic systems (in the sense that the gauge invariant
order parameters in these systems are bosonic), and then we turn
on perturbation, which in our case is the boson-fermion coupling.
We will demonstrate that a non-Fermi liquid can be constructed
based on the unique nature of the strongly interacting bosonic
system.

\section{Expansion of $\epsilon$}


A controlled reliable study of the non-Fermi liquid problem is
generally considered as a very challenging problem, one example of
the difficulties was discussed in Ref.~\cite{sslee}. Over the
years various approximation methods were proposed. We begin by
reviewing the $\epsilon-$expansion developed in
Ref.~\cite{nayak1,nayak2,mrossnfl}, and demonstrate how
perturbation of $\epsilon$ is naturally justified for some
unconventional QCPs. It is often convenient to study interacting
fermions with finite density by expanding at one patch of the
Fermi surface. The low-energy theory of the fermions expanded at
one patch of the fermi surface is \beqn
\mathcal{L}_{f}=\psi^{\dagger}\left(\xi \partial_{\tau} - \ii
v_{F}\partial_{x}-\kappa\partial_{y}^{2}\right)\psi, \eeqn where
$x$ is perpendicular to the fermion surface and $y$ is the tangent
direction. The initial value of $\xi$ is $\xi_0 = 1$, and it will
be renormalized by the fermion self-energy. Our main goal is to
evaluate the fermion self-energy to the leading nontrivial order
of the boson-fermion coupling. We will show that this is
equivalent to the leading nontrivial order of $\epsilon = 1 -
\eta$. At this order of expansion of $\epsilon$, for our purpose
it is sufficient to consider a simple ``effective action" of
$\mathcal{O}(\boldsymbol{x}, \tau)$: \beqn \mathcal{S}_{eff} \sim
\int d^2x d\tau \ \mathcal{O}(\boldsymbol{x}, \tau) (-
\partial_\tau^2 - v^2 \boldsymbol{\nabla}^2)^{1 - \frac{\eta}{2}}
\mathcal{O}(\boldsymbol{x}, \tau) \label{action} \eeqn which will
reproduce the correlation function of $\mathcal{O}(\boldsymbol{x},
\tau)$, assuming we have fully solved the interacting bosonic
system first.

When the boson-fermion coupling is zero, i.e., $g=0$, the system
is at a Gaussian fixed point with the following scaling dimensions
of spacetime coordinates and fields \beqn && [\tau] = -2, \ \ [x]
= -2, \ \ [y] = -1, \cr\cr && [\psi\left(\boldsymbol{x},
\tau\right)] = \frac{3}{2}, \ \ [\mathcal{O}(\boldsymbol{x},
\tau)] = \frac{3}{2} + \frac{\eta}{2} = 2 - \frac{\epsilon}{2}.
\label{scaling} \eeqn We then turn on the boson-fermion
interaction \beqn \int d^2x d\tau \ g \psi^\dagger T \psi
\mathcal{O} \eeqn and consider the perturbative RG at the Gaussian
fixed point. We find that the scaling dimension of $g$ is $[g] =
\epsilon/2$, hence it is weakly relevant if $\epsilon$ is
naturally small, and it may flow to a weakly coupled new fixed
point in the infrared which facilitates perturbative calculations
with expansion of $\epsilon$. Indeed, the beta function of $g^{2}$
at the leading order of $\epsilon$ was derived in
Ref.~\cite{nayak1,nayak2,mrossnfl}:
\begin{equation}
\frac{dg^{2}}{d\log b} = \frac{\epsilon}{2} g^{2} - \Upsilon
g^{4}. \label{eq:beta function of g^2}
\end{equation}
Thus there is a fixed point at weak coupling
$g_{*}^{2}=\epsilon/(2\Upsilon)$, where the parameter $\Upsilon
\sim 1/ (4\pi^{2}v_{F}v)$.



Under the rescaling $x' = x b^{-1}$, namely after integrating out
the short scale degrees of freedom, the fermion acquires a
one-loop self-energy \beqn \delta
\Sigma_{f}\left(\ii\omega,\boldsymbol{p}\right) \sim
g^2 \int d\nu d \boldsymbol{q} \langle
\mathcal{O}^\ast_{\boldsymbol{q}, \nu}
\mathcal{O}_{\boldsymbol{q}, \nu} \rangle G_f(\ii \omega + \ii
\nu,\boldsymbol{q} + \boldsymbol{p} ) \cr\cr \sim g^{2}\int d\nu
dq_{x}\int_{\frac{\Lambda}{\sqrt{b}}}^{\Lambda}dq_{y}
\frac{1}{\left|v^2 q_x^2 +
v^{2}q_{y}^{2}+\omega^{2}\right|^{\frac{1+\epsilon}{2}}} \cr\cr
\times \frac{1}{\ii \left(\omega + \nu\right) - v_{F} \left(p_{x}
+ q_{x} \right) - \kappa\left(p_{y} + q_{y}\right)^{2}}. \eeqn In
the boson correlation function, $v^2 q_x^2$ and $\omega^2$ are
irrelevant compared with $v^2 q_y^2$, hence we first integrate
over $q_x$, and the fermion propagator contributes a factor
$\textrm{sgn}\left(\omega+\nu\right)\ii/ (2v_{F})$. We then
perform the $\nu$ integral and finally integrate $q_y$ over the
momentum shell $\Lambda b^{-1/2}<\left|q_{y}\right|<\Lambda$. The
last integral is evaluated at $\epsilon = 0$, which is valid at
the leading order perturbation of $\epsilon$. This procedure leads
to \begin{equation} \delta \Sigma_{f}\left(\ii
\omega,\boldsymbol{p}\right)= - \ii\omega g^{2}\Upsilon\log b +
O\left(\epsilon^{2}\right).
\end{equation}
Combining the calculations above, at the fixed point $g_\ast^2$,
the renormalized $\ii \xi(\omega) \omega$ in the Fermion Green's
function reads \beqn \ii \xi(\omega) \omega \sim -\ii
\textrm{sgn}\left(\omega\right)\left|\omega\right|^{1-\epsilon/2}.
\label{scale}\eeqn The fermion self-energy, hence the decay rate
of the fermion, scales in the same way as Eq.~\ref{scale}. The
calculation above gives a nearly-marginal non-Fermi liquid
behavior for small but finite $\epsilon$. For small $\eta$ such as
the cases in the Wilson-Fisher fixed points, the calculation of
the scaling of fermion self-energy is not reliable with the
leading order expansion of $\epsilon$ described above.

Here we stress that, our main purpose is to compute $\ii
\xi(\omega)\omega$, or the fermion self-energy to the leading
order of boson-fermion coupling $g^2_\ast \sim \epsilon$, assuming
a weak initial coupling $g$. At higher order expansion of the
boson-fermion coupling, corrections to the boson field self-energy
(for example the standard RPA diagram) from the boson-fermion
coupling needs to be considered. The RPA diagram is proportional
to $\mathcal{L}_{\mathrm{RPA}} \sim |\mathcal{O}_{\omega,
\boldsymbol{q}}|^2 g^2 |\omega|/(v_F \kappa q)$. Several
parameters can be tuned, including the weak coupling fixed point
value of $g^2_\ast$, to make this term weak enough to allow an
energy window where the calculations in this section apply. At the
elementary level, we need the terms in Eq.~\ref{action} to
dominate the RPA effect $|\mathcal{O}_{\omega, \boldsymbol{q}}|^2
g^2 |\omega|/(v_F \kappa q)$. A field $\mathcal{O}$ at momentum
$\boldsymbol{q}$ should correspond to energy scale $\omega \sim v
q$. For Eq.~\ref{action} at $\eta = 1$ to dominate the RPA effect,
we need $q > g^2 / (v_F \kappa)$, or $\omega > g^2 v / (v_F
\kappa)$. If we start with a weak initial bare coupling constant
$g_0$, and also $\epsilon \ll 1$ hence the fixed point value of
$g_\ast$ is also perturbative, there is a sufficiently large
energy window for our result. Tuning the parameter $v/v_F$ and
$\kappa$ can further expand the energy window. A full analysis of
the term $\mathcal{L}_{\mathrm{RPA}} \sim |\mathcal{O}_{\omega,
\boldsymbol{q}}|^2 g^2 |\omega|/(v_F \kappa q)$ in the bosonic
sector of the theory in the infrared limit requires more detailed
analysis because $\mathcal{O}_{\omega, \boldsymbol{q}}$ is a
composite operator in the field theories discussed in the next
section.

\section{Candidate unconventional QCPs}


{\it (1) Bosonic-QED-Chern-Simons theory}

In the following we will discuss candidate QCPs which suffice the
desired condition $\eta \sim 1$, or $\epsilon \ll 1$. When we
study the pure bosonic sector of the theory, we ignore the
coupling to the fermions, assuming the boson-fermion coupling is
weak, which is self-consistent with the conclusion in the previous
review section that the boson-fermion interaction will flow to a
weakly coupled fixed point $g_\ast^2 \sim \epsilon$. As we stated
in the previous section, we will start with a weak boson-fermion
coupling $g$, and eventually we only compute the fermion
self-energy to the leading nontrivial order of the fixed point
$g^2_\ast \sim \epsilon$. In the purely bosonic theory, the
scaling of the space-time has the standard Lorentz invariance. To
avoid confusion, we use ``$[ \ ]$" to represent scaling dimensions
under the scaling Eq.~\ref{scaling} of the one-patch theory in the
previous section, and ``$\{ \ \}$" represent the scaling dimension
in the Lorentz invariant purely bosonic theory. At a QCP, multiple
operators will become ``critical", namely multiple operators can
have power-law correlation. We will demand that the operator with
the strongest correlation (smallest scaling dimension) satisfy the
desired condition, since this is the operator that provides the
strongest scattering with the electrons.

We consider $(2+1)d$ bosonic quantum electrodynamics (QED) with
$N$ flavors of bosons coupled to a noncompact $\U(1)$ gauge field
with a Chern-Simons term: \beqn \mathcal{L}_\mathrm{bQED} &=&
\sum_{\alpha = 1}^2 \sum_{a = 1}^{N/2} |(\partial_\mu - \ii
b_\mu)z_{\alpha, a}|^2 + r (z^\dagger_{\alpha, a} z_{\alpha, a})
\cr\cr &+& u ( \sum_{\alpha, a} |z_{\alpha, a}|^2 )^2 + u'
\sum_{\alpha = 1}^2 ( \sum_{a = 1}^{N/2} |z_{\alpha, a}|^2 )^2
\cr\cr &+& \frac{\ii kN}{4\pi} b \wedge d b. \label{lqed}\eeqn The
following operators are gauge invariant composite fields, which we
assume are all at zero momentum: \beqn \mathcal{O}_0 =
\sum_{\alpha = 1}^2 \sum_{a = 1}^{N/2} z^\dagger_{\alpha, a}
z_{\alpha, a}, \ \ \ \mathcal{O}_{1,3} = \sum_{a = 1}^{N/2}
z^\dagger_{a} \sigma^{1,3} z_{a}. \label{field}\eeqn Potential
applications of this field theory to strongly correlated systems
will be discussed later.

\begin{figure}
\centering
\includegraphics[width=\linewidth]{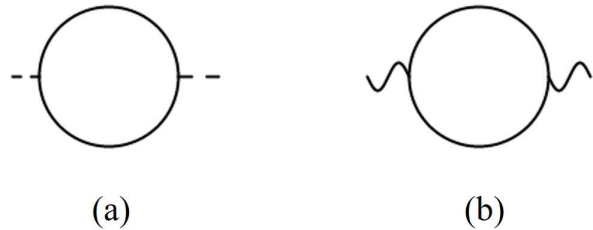}
\caption{The self-energy of field $\sigma_+$ and gauge field
$b_\mu$ in the large$-N$ limit.} \label{loop}
\end{figure}

To compute their scaling dimensions, we introduce two
Hubbard-Stratonovich(HS) fields to decouple the quartic
potentials: \beqn \mathcal{L}'_\mathrm{bQED} &=& \sum_{\alpha =
1}^2 \sum_{a = 1}^{N/2} |(\partial_\mu - \ii b_\mu)z_{\alpha,
a}|^2 + r (z^\dagger_{\alpha, a} z_{\alpha, a}) \cr\cr &+& \ii
\sigma_+ \mathcal{O}_0 + \ii \sigma_- \mathcal{O}_3 + \frac{1}{2u'
+ 4u}\sigma_+^2 + \frac{1}{2u'}\sigma_-^2 \cr\cr &+& \frac{\ii
kN}{4\pi} b \wedge d b. \label{lqed2} \eeqn We will consider the
following two scenarios: (1) $u' \to 0, u > 0$, where $\sigma_-$
is fully suppressed and the system has a full $\SU(N) \times
\U(1)_T$ symmetry, where the $\U(1)_T$ is the ``topological
symmetry" that corresponds to the conservation of the gauge flux;
and (2) $u,u' > 0$ when the $\SU(N)$ symmetry is broken down to
$\SU(N/2) \times \SU(N/2) \times \U(1) \rtimes Z_2$, where the
$\U(1)\rtimes Z_2$ is the symmetry within the Pauli matrix space
in Eq.~\ref{field}.

In scenario (1) with a full $\SU(N)$ symmetry, at the critical
point $r = 0$, the field $\sigma_+$ acquires a self-energy in the
large$-N$ limit \beqn \Sigma_{\sigma_+} (p) =
N\int\frac{d^3q}{(2\pi)^3}\frac{1}{q^2(q+p)^2} = \frac{N}{8p}.
\eeqn Hence the propagator of field $\sigma_+$ in the large$-N$
limit reads \beqn G_{\sigma_+}(p) = 1/\Sigma_{\sigma_+} =
\frac{8p}{N}.\eeqn

Similarly, for the gauge field, the self-energy in the large$-N$
limit is \beqn \Sigma_{b,\mu\nu}(p)
&=&-N\int\frac{d^3q}{(2\pi)^3}\frac{(2q+p)_\mu(2q+p)_\nu}{q^2(q+p)^2}\cr\cr
&=&\frac{N}{16p}(p^2\delta_{\mu\nu}-p_\mu p_\nu). \eeqn When
combined with the Chern-Simons term, in the Landau gauge, the
gauge field has the following large$-N$ propagator~\cite{wen-wu}
\beqn
G_{b,\mu\nu}(p)=\frac{1}{Np}\left(F\left(\delta_{\mu\nu}-\frac{p_\mu
p_\nu}{p^2}\right)+H\frac{\epsilon_{\mu\nu\rho}p^\rho}{p}\right),\eeqn
where \beqn F=\frac{16\pi^2}{\pi^2+64k^2}, \ \ \ H=-\frac{128\pi
k}{\pi^2+64k^2}. \eeqn

\begin{figure}
\centering
\includegraphics[width=\linewidth]{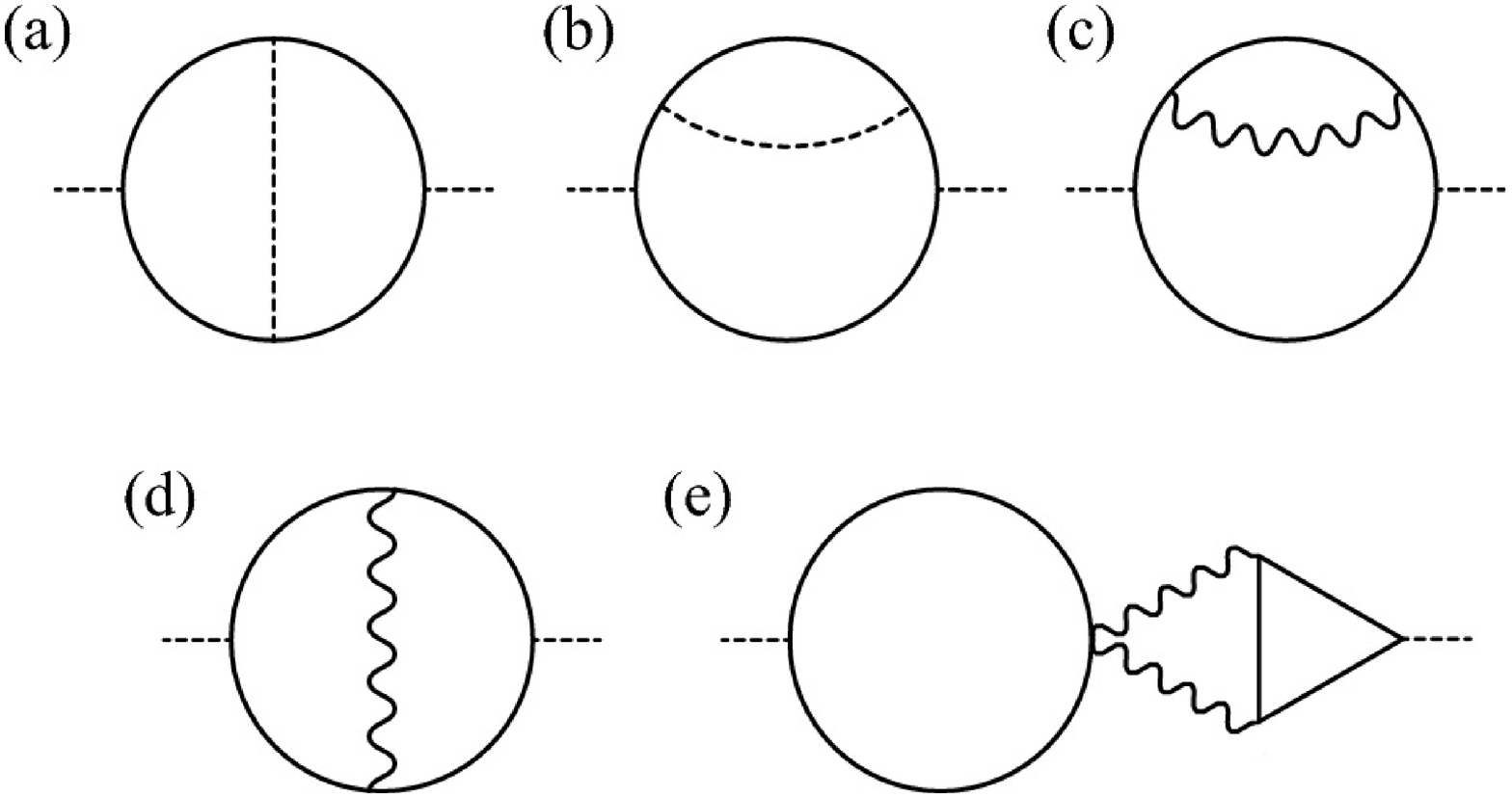}
\caption{In scenario (1), diagrams $(a)-(e)$ contribute to the
anomalous dimension of $\mathcal{O}_0$ in Eq.~\ref{lqed} or
equivalently $\sigma_+$ in Eq.~\ref{lqed2}; while only diagrams
$(a)-(d)$ contribute to the anomalous dimension of
$\mathcal{O}_{1,3}$. The solid line represents the propagator of
$z_{\alpha,a}$, the dashed and wavy lines represent the large$-N$
propagators of $\sigma_{+}$ and $b_\mu$ respectively.}
\label{feynman} \end{figure}

After introducing the HS fields, the scaling dimension of the
composite operator $\mathcal{O}_{0}$ of the original field theory
Eq.~\ref{lqed} is ``transferred" to the scaling dimension of the
HS fields $\sigma_{+}$. To the order of $O(1/N)$, the Feynman
diagrams in \figref{feynman} contribute to the $\sigma_+$ self
energy, which was computed in Ref.~\cite{wen-wu}.

But it is evident that in the large$-N$ limit, the scaling
dimension of $\sigma_+$ (and the scaling dimension of operator
$\mathcal{O}_{0}$ of the original field theory Eq.~\ref{lqed}) is
$\lim_{N \rightarrow \infty}\{\mathcal{O}_{0}\} = 2$, hence it
does not meet the desired condition. When $\mathcal{O}_{0}$
couples to the Fermi surface, the boson-fermion coupling will be
irrelevant in the one patch theory discussed in the previous
section according to the scaling of space-time Eq.~\ref{scaling}.


The scaling dimension of $\sigma_{1,3}$ equal to each other with a
full $\SU(N)$ symmetry, and unlike $\mathcal{O}_0$, they have
scaling dimension $1$ in the large$-N$ limit. The $1/N$
corrections to their anomalous dimensions come from diagram
$(a)-(d)$ in Fig.~\ref{feynman}, or equivalently through the
standard momentum shell RG: \beqn \{ \mathcal{O}_{1,3} \} = 1 +
\frac{16}{3\pi^2 N} - \frac{4}{3\pi^2 N}F. \eeqn
Ref.~\cite{kaulsachdev} and references therein have computed
scaling dimensions of gauge invariant operators for theories with
matter fields coupled with a $\U(1)$ gauge field, without a
Chern-Simons term. Our result is consistent with these previous
references, since $\lim_{k \rightarrow 0} \{\mathcal{O}_{1,3}\} =
1 - 16/( \pi^2 N)$, which is the result of the CP$^{N-1}$ model
with a noncompact gauge field. Also, in the limit of $k
\rightarrow + \infty$, our result is consistent with
Ref.~\cite{kaulsachdev} when the fermion component is taken to be
infinity, since both limits suppress the gauge field fluctuation
completely. In general operators $\mathcal{O}_{1,3}$ have stronger
correlations than $\mathcal{O}_{0}$, hence they will make stronger
contributions to scattering when coupled with the fermi surface.
As an example, the anomalous dimension of $\mathcal{O}_{1,3}$ with
$k = 1/2$ reads \beqn \eta_{1,3} \sim 1 - \frac{0.57}{N}, \eeqn
which is reasonably close to $1$ even for the most physically
relevant case with $N = 2$.

In scenario (2) we should keep both $\sigma_+$ and $\sigma_-$ in
the calculation, and both $\sigma_{\pm}$ (operator
$\mathcal{O}_{0}$ and $\mathcal{O}_{3}$ in theory Eq.~\ref{lqed})
have scaling dimension 2 in the large$-N$ limit~\cite{B2019}. Now
$\mathcal{O}_{1}$ has the strongest correlation, and at the order
of $O(1/N)$, its scaling dimension reads: \beqn \{ \mathcal{O}_1
\} &=& 1 + \frac{8}{3\pi^2 N} - \frac{4}{3\pi^2 N}F. \eeqn When $k
= 1$, its anomalous dimension reads \beqn \eta_1 \sim 1 -
\frac{0.037}{N}, \eeqn which is always very close to $1$. Using
the formalism reviewed in the previous section, by coupling to
$\mathcal{O}_{1}$, the fermion self-energy would scale as
$\Sigma_{f}\left(\ii \omega, \boldsymbol{p}\right) \sim-\ii
\textrm{sgn}\left(\omega\right)\left|\omega\right|^{0.99}$ for
$N=2$.

The field theory Eq.~\ref{lqed} describes a quantum phase
transition from a topological order with Abelian anyons to an
ordered phase that spontaneously breaks the global flavor
symmetry. The flavor symmetry can be either a full $\SU(N)$
symmetry (scenario {\it 1}) or $\SU(N/2) \times \SU(N/2) \times
\U(1) \rtimes Z_2$ (scenario {\it 2}). So far we have assumed that
the gauge invariant $\mathcal{O}_{1,3}$ have zero momentum, hence
they cannot be the ordinary antiferromagnetic N\'{e}el order
parameter. They must be translational invariant order parameters
with nontrivial representation under the internal symmetry group,
for example they could be the quantum spin Hall order parameter
for $N = 2$.

The topological order described by the Chern-Simons theory with $N
= 2$, $k = 1$ is the most studied state in condensed matter
theory. This topological order is the $\U(1)_2$ or equivalently
the $\SU(2)_1$ topological order with semionic anyons. It is the
most natural topological order that can be constructed from the
slave particle formalism~\cite{wen2002}. And recently it was
conjectured that this topological order is also related to the
parent state of the cuprates high temperature
superconductor~\cite{xuhall} motivated by the giant thermal Hall
signal observed~\cite{chall}.

Another interesting scenario is when $N = 2$, $k = 0$ and $u > 0$.
In this case Eq.~\ref{lqed} is the same field theory as the
easy-plane deconfined QCP between the inplane antiferromagnetic
N\'{e}el order and the valence bond solid state on the square
lattice. Recent numerical studies have shown that this quantum
phase transition may be continuous, and the scaling dimension of
both $\mathcal{O}_{0}$ and $\mathcal{O}_{3}$ are fairly close to
$1$ based on numerical
results~\cite{deconfinedualnumeric,karthik}. It has been proposed
that this field theory is self-dual~\cite{ashvinlesik}, and it is
dual to the transition between the bosonic symmetry protected
topological (SPT) phase and the trivial
phase~\cite{deconfinedual,potterdual}, which is directly describe
by a noncompact QED with $N = 2$ flavors of Dirac fermion matter
fields~\cite{Tarun_PRB2013,lulee}. The tuning parameter for this
topological transition is instead coupled to $\mathcal{O}_{3}$.
Hence this SPT-trivial transition is also a candidate quantum
phase transition which meets the desired criterion proposed in our
paper that leads to a nearly-marginal fermi liquid. But in these
cases there are other fields (for example the inplane N\'{e}el
order parameter) with smaller scaling dimensions, and we need to
assume that these operators carry finite lattice momentum hence
couple to the Fermi surface differently.

{\it (2) Gross-Neveu-Yukawa QCP}

Another candidate QCP that likely suffices the desired condition
$\eta\sim 1$ is the Gross-Neveu-Yukawa QCP with $N-$flavors of
Dirac fermion: \beqn \mathcal{L}_\mathrm{GNY} &=& \sum_{a = 1}^N
\bar{\chi}_a \gamma_\mu \partial_\mu \chi_a + g \phi \bar{\chi}_a
\chi_a \cr\cr &+& (\partial \phi)^2 + r \phi^2 + u \phi^4.
\label{GN} \eeqn At the critical point $r = 0$, both $u$ and $g$
flows to a fixed point. In our context, the QCP describes a
bosonic or spin system, hence $\chi$ is viewed as a fermionic
slave particle of spin, $i.e.$ the spinon, and we assume that
$\chi$ is coupled to a $Z_2$ gauge field, namely the system is a
$Z_2$ spin liquid with fermionic spinons. But the dynamical $Z_2$
gauge field does not lead to extra singular corrections to low
energy correlation functions of gauge invariant operators, hence
the universality class of Eq.~\ref{GN} is still identical to the
Gross-Neveu-Yukawa (GNY) theory, as long as we only focus on gauge
invariant operators.

The GNY QCP can still be solved in the large$-N$ limit, and the
cases with finite $N$ can approached through a $1/N$ expansion. At
the GNY QCP coupled with a $Z_2$ gauge field, the gauge invariant
operator with the lowest scaling dimension is $\phi$, and its
scaling dimension can be found in Ref.~\cite{GNN} and references
therein: \beqn \{ \phi \} \sim 1 - \frac{16}{3\pi^2 N}. \eeqn
Other gauge invariant operators such as $\bar{\chi} T \chi$ with a
$\SU(N)$ matrix $T$ have much larger scaling dimension at the GNY
QCP, for example $\{ \bar{\chi} T \chi \} = 2$ in the large$-N$
limit. If we replace the $Z_2$ gauge field by a $\U(1)$ gauge
field, the $\U(1)$ gauge fluctuation will enhance the correlation
of $\phi$, hence increases $\epsilon = 1 - \eta$ compared with the
situation with only a $Z_2$ gauge field. Hence a GNY QCP with a
$\U(1)$ gauge field is less desirable according to our criterion.

The GNY QCP coupled with a $Z_2$ gauge field can be realized in
various lattice model Hamiltonians for quantum antiferromagnet.
For example, for $\SU(M)$ spin systems on the triangular lattice
with a self-conjugate representation on each site, using the
fermionic spinon formalism, when there is a $\pi-$flux through
half of the triangles, there are $N = 2M$ components of Dirac
fermions at low energy~\cite{luz2}. $\SU(M)$ quantum magnet may be
realized in transition metal oxides with orbital
degeneracies~\cite{SU41,SU42,SU43}, and also cold atom systems
with large hyperfine spins~\cite{wu1,wu2,wu3,xusun}. Recently it
was also proposed that an approximate $\SU(4)$ quantum
antiferromagnet can be realized in some of the recently discovered
Moir\'{e} systems~\cite{xuleon,xuFM,fumag}, and a $\SU(4)$ quantum
antiferromagnet on the triangular lattice may realize the
$Z_2-$gauged GNY QCP with $N = 8$ (with lower spatial symmetry
compared with $\SU(2)$ systems as was pointed out in
Ref.~\cite{zhangmao}). On the other hand, a $\SU(M)$ spin systems
on the honeycomb lattice can potentially realize the GNY QCP with
$N = 2M$ (with zero flux through the hexagon) or $N = 4M$ (with
$\pi-$flux through the hexagon).

The operator $\phi$ is odd under time-reversal and spatial
reflection, hence physically $\phi$ corresponds to the spin
chirality order. Hence the $Z_2-$gauged GNY QCP is a quantum phase
transition between a massless spin liquid and a chiral spin
liquid.

Non-Fermi liquid is often observed only at a finite
temperature/energy window in experiments. At the infrared limit,
the non-Fermi liquid is usually preempted by other instabilities,
for example a dome of
superconductor~\cite{nflpair1,nflpair2,nflpair3}. In
Ref.~\cite{nflpair1} the instability of non-Fermi liquid towards
the superconductor dome was systematically studied in the
framework of the $\epsilon-$expansion. According to
Ref.~\cite{nflpair1}, when $\mathcal{O}$ is an order parameter at
zero momentum, at $\epsilon = 0$ the superconductor instability
will occur at an exponentially suppressed temperature/energy scale
$\Delta_{\mathrm{sc}} \sim \Lambda_{\omega} \exp(- A / |g_0|) $,
where $g_0$ is the bare boson-fermion coupling constant. In our
case the estimate of the superconductor instability is complicated
by the fact that $\mathcal{O}$ is a composite field, but the
qualitative exponentially-suppressed form of
$\Delta_{\mathrm{sc}}$ is not expected to change because $g$ is
still at most a marginally relevant coupling. When $\epsilon = 0$,
the imaginary part of the fermi self-energy (the inverse of
quasi-particle life-time) scales linearly with $\omega$. Because
the bare electron dispersion has no imaginary part at all, the
imaginary part of the self-energy should be much easier to observe
compared with the real part, assuming other scattering mechanisms
of the fermions are weak enough. The scaling behavior of the
fermion self-energy is also observable numerically like
Ref.~\onlinecite{meng23}. This linear scaling behavior of the
imaginary part of self-energy is observable for fermionic
excitations at energy scale $\omega > \Delta_{\mathrm{sc}}$, .
Hence above the superconductor energy scale
$\Delta_{\mathrm{sc}}$, the non-Fermi liquid behavior is
observable. This result should still hold for small enough
$\epsilon$. \footnote{In Ref.~\cite{nflpair1}, the non-Fermi
liquid energy scale $E_{\mathrm{nfl}}$ is defined as the energy
scale where the fermi velocity $v_F$ is renormalized strongly from
its bare value, hence $E_{\mathrm{nfl}}$ was defined based on the
real part of the fermion self-energy. In other words the
$E_{\mathrm{nfl}}$ was defined as the scale where the real part of
self-energy dominates the bare energy in the Green's function. But
since the bare dispersion of fermion is difficult to observe, and
the bare fermion energy has no imaginary part at all, we prefer to
use the imaginary part of fermion self-energy as a characteristic
definition of non-Fermi liquid state.}

\section{Conclusion}


In this work we proposed a mechanism based on which a nearly
marginal non-fermi liquid can be constructed with a controlled
method in an energy window. This mechanism demonstrates that two
exceptional phenomena beyond the standard Landau's paradigm,
$i.e.$ the non-Landau quantum phase transitions and the non-fermi
liquid may be connected: a non-Landau quantum phase transition can
have a large anomalous dimension $\eta \sim 1$, which physically
justifies and facilitates a perturbative calculation of the
Boson-Fermion coupling fixed point. Several candidate QCPs that
suffice this condition were proposed, including topological
transitions from Abelian topological orders to an ordered phase,
and a Gross-Neveu-Yukawa transition of $Z_2$ spin liquids.

We would like to compare our construction of non-fermi liquid
states and the constructions based on the SYK related models. In
the constructions based on SYK-like models, the existence of a
strange-metal like phase was based on the fact that in the soluble
limit, $i.e.$ in the SYK model the scaling dimension of fermion is
$1/4$ (scaling with time only). But since the definition of the
electric current operator in these constructions is proportional
to the perturbation away from the SYK model, the current-current
correlation function and the electrical conductivity is small in
the energy window where the construction applies. Recently an
improved construction was proposed which can produce the Planckian
metal observed in cuprates materials~\cite{sachdevplanckian}. In
our construction, since the boson-fermion coupling will flow to a
weakly coupled fixed point, the scattering rate of the fermion due
to the boson-fermion coupling is expected to be low. We will
further study if a Planckian metal like state can be constructed
by developing our current approach. In this future exploration, a
mechanism of momentum relaxation, for instance the disorder, or
Umklapp process, needs to be introduced.

This work is supported by NSF Grant No. DMR-1920434, the David and
Lucile Packard Foundation, and the Simons Foundation.

\bibliography{mfl}

\end{document}